\documentclass[runningheads]{llncs}

\usepackage{amsmath}
\usepackage{multirow}
\usepackage[table,xcdraw]{xcolor}
\usepackage{graphicx}
\usepackage{textcomp}
\usepackage{subfig}
\usepackage[utf8x]{inputenc}
\usepackage{csquotes}
\usepackage{flexisym}
\newcommand{\ignore}[1]{}
\usepackage[hidelinks]{hyperref}
\usepackage{hyphenat}
\usepackage{wrapfig}

\usepackage[ruled,vlined,linesnumbered]{algorithm2e}

\usepackage{amssymb}
\usepackage[bottom,multiple]{footmisc}
\usepackage{wrapfig}
\usepackage{nicefrac}
\usepackage{bm}
\usepackage{mathrsfs}

\usepackage[font=scriptsize]{caption}

\newcommand{\pactivities}{\mathcal{A}}

\newcommand{\presources}{\mathcal{R}}

\newcommand{\ptimes}{\mathcal{T}}

\newcommand{\multiset}{\mathcal{B}}

\newcommand{\universe}{\mathcal{U}}

\newtheorem{exmp}{Example}

\newcommand{\caseuniverse}{\mathcal{C}}
\newcommand{\eventuniverse}{\mathcal{E}}
\newcommand{\attuniverse}{\mathcal{N}}
\newcommand{\valueuniverse}{\mathcal{U}}

\DeclareMathSymbol{\mlq}{\mathord}{operators}{'134}
\DeclareMathSymbol{\mrq}{\mathord}{operators}{'42}

\usepackage{mathtools}

\begin{document}
\title{Privacy-Preserving Data Publishing in Process Mining}
\titlerunning{Privacy-Preserving Data Publishing in Process Mining}
%
%
\author{Majid Rafiei\orcidID{0000-0001-7161-6927} \and
	Wil M.P. van der Aalst\orcidID{0000-0002-0955-6940}}
\authorrunning{Majid Rafiei and Wil M.P. vand der Aalst}
%
\institute{Chair of Process and Data Science, RWTH Aachen University, Aachen, Germany \\
 }
\maketitle              

\begin{abstract}

Process mining aims to provide insights into the actual processes based on event data. These data are often recorded by information systems and are widely available. However, they often contain sensitive private information that should be analyzed responsibly. Therefore, privacy issues in process mining are recently receiving more attention. Privacy preservation techniques obviously need to modify the original data, yet, at the same time, they are supposed to preserve the data utility. 
Privacy-preserving transformations of the data may lead to incorrect or misleading analysis results. Hence, new infrastructures need to be designed for publishing the privacy-aware event data whose aim is to provide metadata regarding the privacy-related transformations on event data without revealing details of privacy preservation techniques or the protected information. In this paper, we provide formal definitions for the main anonymization operations, used by privacy models in process mining. These are used to create an infrastructure for recording the privacy metadata. We advocate the proposed privacy metadata in practice by designing a privacy extension for the XES standard and a general data structure for event data which are not in the form of standard event logs.


\end{abstract}
\section{Introduction}\label{sec:introduction}

No one doubts that data are extremely important for people and organizations and their importance is growing. Hence, the interest in \textit{data science} is rapidly growing. Of particular interest are the so-called event data used by the \textit{process mining} techniques to analyze end-to-end processes. Process mining bridges the gap between traditional model-based process analysis, e.g., simulation, and data-enteric analysis, e.g., data mining. It provides fact-based insights into the actual processes using event logs \cite{van2016process}. The three basic types of process mining are: \textit{process discovery}, \textit{conformance checking}, and \textit{enhancement} \cite{van2016process}. 


An event log is a collection of events, and each event is described by its attributes. The main attributes required for process mining are \textit{case id}, \textit{activity}, \textit{timestamp}, and \textit{resource}. \autoref{tbl:sample_evenlog} shows an event log recorded by an information system in a hospital, where $\bot$ indicates that the corresponding attribute was not recorded. Some of the event attributes may refer to individuals. For example, in the health-care context, the \textit{case id} may refer to the patients whose data are recorded, and the \textit{resource} may refer to the employees performing activities for the patients, e.g., nurses or surgeons. When the individuals' data are explicitly or implicitly included, \textit{privacy issues} arise. According to regulations such as the European General Data Protection Regulation (GDPR) \cite{voss2016european}, organizations are compelled to consider the privacy of individuals while analyzing their data.


The \textit{privacy} and \textit{confidentiality} issues in process mining are recently receiving more attention \cite{rafieiWA18,rafiei2019role,pretsaICPM2019,MannhardtKBWM19}. The proposed methods cover a range of solutions from privacy/confidentiality frameworks to privacy guarantees. Privacy preservation techniques often apply some anonymization operations to modify the data in order to fulfill desired \textit{privacy requirements}, yet, at the same time, they are supposed to preserve the \textit{data utility}. The transformed event log may only be suitable for specific analyses. For example, in \cite{rafiei2019role}, the privacy requirement is to \textit{discover social networks of the individuals (resources) involved in a process without revealing their activities}, and the resulting event log is only suitable for the \textit{social network discovery}.  
Moreover, the original event log may be transformed to another form of event data which does not have the structure of a standard event log. For example, in \cite{rafieiWA18}, the privacy requirement is to \textit{discover processes without revealing the sequence of activities performed for the individuals (cases)}, where the transformed event data are not in the form of event logs and contain only \textit{directly follows relations between activities}, which is merely suitable for the \textit{process discovery}. Therefore, the modifications made by privacy preservation techniques need to be reflected in the transformed event data to inform the data analysts. 

In this paper, for the first time, we formalize the main anonymization operations on the event logs and exploit them as the basis of an infrastructure for proposing privacy metadata, we also design a privacy extension for the XES standard and a general data structure to cope with the event data generated by some privacy preservation techniques which are not in the form of an event log. The proposed metadata, along with the provided tools support, supply privacy-aware event data publishing while avoiding inappropriate or incorrect analyses.

The remainder of the paper is organized as follows. In Section~\ref{sec:motivation}, we explain the motivation of this research. Section~\ref{sec:related_work} outlines related work. In Section~\ref{sec:prelimineries}, formal models for event logs are presented. We explain the privacy-preserving data publishing in process mining in Section~\ref{sec:ppdp}, where we formalize the main anonymization operations, privacy metadata are proposed, and the tools support is presented. Section~\ref{sec:conclusions} concludes the paper.

\begin{table}[tb]
	\centering
	\begin{minipage}{.49\textwidth}
		\centering
		\tiny
			\caption{An event log (each row represents an event).}\label{tbl:sample_evenlog}
	\begin{tabular}{|l|l|l|l|l|l|}
		\hline
		Case Id & Act.             & Timestamp        & Res.  & Age & Disease     \\ \hline
		1       & a    & 01.01.2019-08:30:10 & E1 & 22  & Flu         \\ 
		1       & b            & 01.01.2019-08:45:00 & D1   & 22  & Flu         \\ 
		2       & a    & 01.01.2019-08:46:15 & E1 & 30  & Infection   \\ 
		3       & a    & 01.01.2019-08:50:01 & E1 & 32  & Infection   \\ 
		4       & a    & 01.01.2019-08:55:00 & $\bot$ & 29  & Poisoning   \\ 
		1       & e         & 01.01.2019-08:58:15 & E2 & 22  & Flu         \\ 
		4       & b            & 01.01.2019-09:10:00 & D2   & 29  & Poisoning   \\ 
		4       & r        & 01.01.2019-09:30:00 & B1    & 29  & Poisoning   \\ 
		2       & d & 01.01.2019-09:46:00 & E3 & 30  & Infection   \\ 
		3       & d & 01.01.2019-10:00:25 & E3 & 32  & Infection   \\ 
		2       & f      & 01.01.2019-10:00:05 & N1    & 30  & Infection   \\  
		3       & f      & 01.01.2019-10:15:22 & N1    & 32  & Infection   \\ 
		4       & e         & 01.01.2019-10:30:35 & E2 & 29  & Poisoning   \\ 
		2       & f      & 01.02.2019-08:00:45 & N1    & 30  & Infection   \\ 
		2       & b            & 01.02.2019-10:00:00 & D2   & 30  & Infection   \\ 
		3       & b            & 01.02.2019-10:15:30 & D1   & 32  & Infection   \\ 
		2       & e         & 01.02.2019-14:00:00 & E2 & 30  & Infection   \\ 
		3       & e         & 01.02.2019-14:15:00 & E2 & 32  & Infection   \\  \hline
	\end{tabular}
	\end{minipage}
	\begin{minipage}{.49\textwidth}
		\centering
		\tiny
		\caption{An anonymized event log (each row represents an event).}\label{tbl:sample_evenlog_anon}
		\begin{tabular}{|l|l|l|l|l|l|}
			\hline
			Case Id & Act.             & Timestamp        & Res.  & Age & Disease     \\ \hline
			1       & a    & 01.01.2019-08:30:00 & E1 & 22  & Flu         \\ 
			1       & b            & 01.01.2019-08:45:00 & D1   & 22  & Flu         \\ 
			2       & a    & 01.01.2019-08:46:00 & E1 & 30  & Infection   \\ 
			3       & a    & 01.01.2019-08:50:00 & E1 & 32  & Infection   \\ 
			4       & a    & 01.01.2019-08:55:00 & $\bot$ & 29  & Poisoning   \\ 
			1       & e         & 01.01.2019-08:58:00 & E2 & 22  & Flu         \\ 
			4       & b            & 01.01.2019-09:10:00 & D2   & 29  & Poisoning   \\ 
			4       & r        & 01.01.2019-09:30:00 & $\bot$    & 29  & Poisoning   \\ 
			2       & d & 01.01.2019-09:46:00 & E3 & 30  & Infection   \\ 
			3       & d & 01.01.2019-10:00:00 & E3 & 32  & Infection   \\ 
			2       & g      & 01.01.2019-10:00:00 & N1    & 30  & Infection   \\ 
			3       & g      & 01.01.2019-10:15:00 & N1    & 32  & Infection   \\ 
			4       & e         & 01.01.2019-10:30:00 & E2 & 29  & Poisoning   \\ 
			2       & k      & 01.02.2019-08:00:00 & N1    & 30  & Infection   \\ 
			2       & b            & 01.02.2019-10:00:00 & D2   & 30  & Infection   \\ 
			3       & b            & 01.02.2019-10:15:00 & D1   & 32  & Infection   \\ 
			2       & e         & 01.02.2019-14:00:00 & E2 & 30  & Infection   \\ 
			3       & e         & 01.02.2019-14:15:00 & E2 & 32  & Infection   \\  \hline
		\end{tabular}
	\end{minipage}
	\label{fig:exp-tts-corr}
\end{table} 

\section{Motivation}\label{sec:motivation}
Compare \autoref{tbl:sample_evenlog_anon} with \autoref{tbl:sample_evenlog}, they both look like an original event log containing all the main attributes to apply process mining techniques. However, \autoref{tbl:sample_evenlog_anon} is derived from \autoref{tbl:sample_evenlog} by randomly substituting some activities (f was substituted with g and k), generalizing the timestamps (the timestamps got generalized to the \textit{minutes} level), and suppressing some resources (B1 was suppressed). Hence, a performance analysis based on \autoref{tbl:sample_evenlog_anon} may not be as accurate as the original event log, the process model discovered from \autoref{tbl:sample_evenlog_anon} contains some fake activities, and the social network of resources is incomplete.
The main motivation of this research is to provide concrete privacy metadata for process mining without exposing details of privacy/confidentiality techniques or the protected sensitive information so that the analysts are aware of the changes and avoid inappropriate analyses. 

\begin{figure}[bt]
	\centering
	\frame{\includegraphics[width=0.90\textwidth]{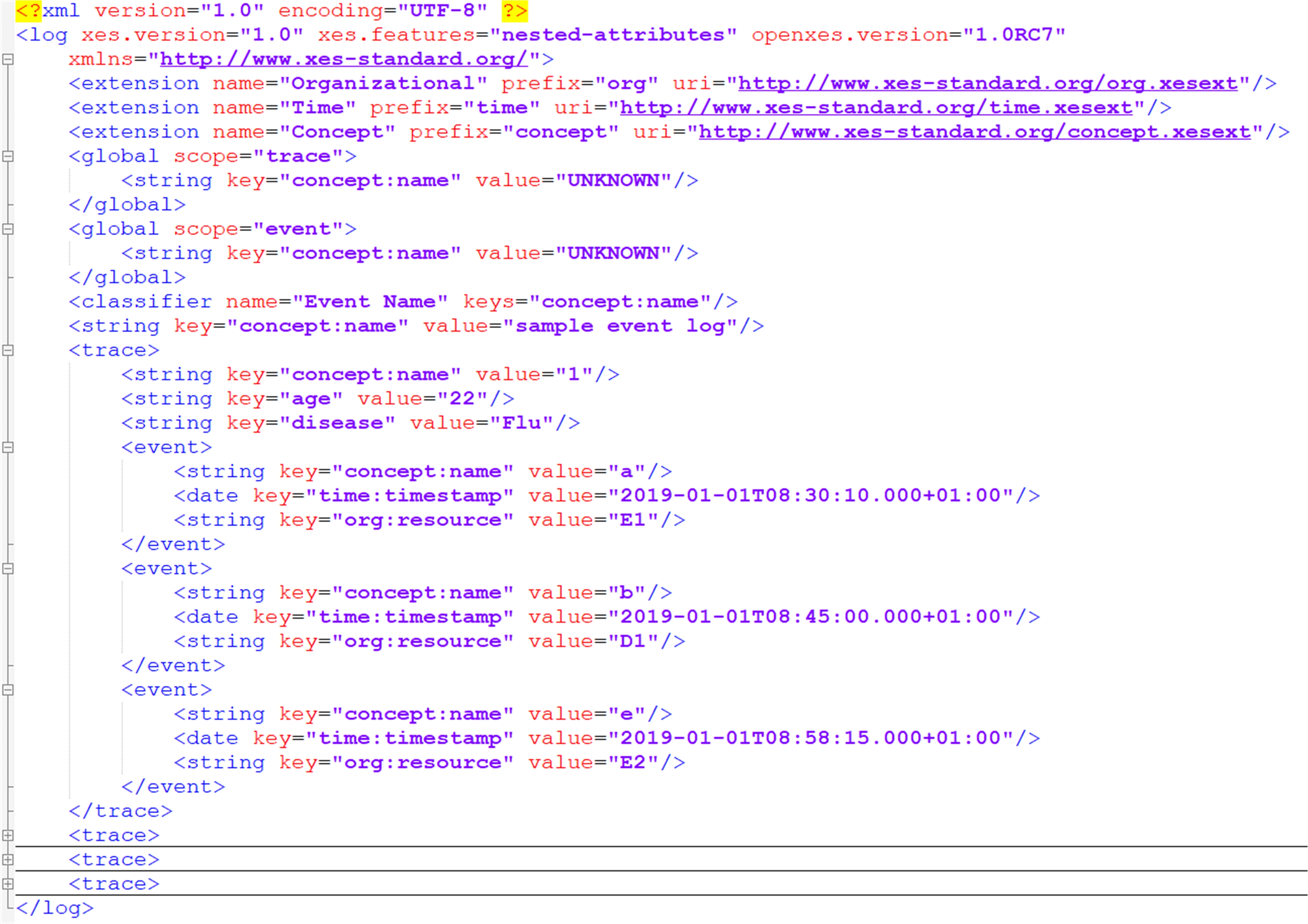}}
	\caption{The XES format for the event log \autoref{tbl:sample_evenlog}, showing only the first case (trace). In XES, the log contains traces and each trace contains events. The log, traces, and events have attributes, and \textit{extensions} may define new attributes. The log declares the extensions used in it. The \textit{global attributes} are the ones that are declared to be mandatory with a default value. The \textit{classifiers} assign identity to each event, which makes it comparable to the other events.}\label{fig:xes}
\end{figure}

Process mining benefits from a well-developed theoretical and practical foundation letting us perform this research.  
In theory, event logs, as the input data types, have a concrete structure by the definition. In practice, the IEEE Standard for eXtensible Event Stream (XES)\footnote{\scriptsize http://www.xes-standard.org/} is defined as a grammar for a tag-based language whose aim is to provide a unified and extensible methodology for capturing systems behaviors by means of event logs, e.g., \autoref{fig:xes} shows the first case of the event log \autoref{tbl:sample_evenlog} in the XES format. 
In this paper, the XES standard will be used to show the concrete relation between the theory and practice, but the concepts are general.


\section{Related Work}\label{sec:related_work}
In process mining, the research field of confidentiality and privacy received rather little attention, although the \textit{Process Mining Manifesto} \cite{van2011process} already pointed out the importance of privacy. In \cite{van2016responsible}, \textit{Responsible Process Mining} (RPM) is introduced as the sub-discipline focusing on possible negative side-effects of applying process mining. 
In \cite{mannhardt2018privacy}, the aim is to provide an overview of privacy challenges in process mining in human-centered industrial environments. 
In \cite{burattin2015toward}, a possible approach toward a solution, allowing the outsourcing of process mining while ensuring the confidentiality of dataset and processes, has been presented. 
In \cite{MichaelKMBR19}, the authors propose a privacy-preserving system design for process mining, where a user-centered view is considered to track personal data. In \cite{rafieiWA18,rafieiWA19}, a framework is introduced, which provides a generic scheme for confidentiality in process mining. In \cite{rafiei2019role}, the aim is to provide a privacy-preserving method for discovering roles from event data, which can also be exploited for generalizing \textit{resources} as individuals in event logs.    
In \cite{liu2016towards}, the authors consider a cross-organizational process discovery context, where public process model fragments are shared as safe intermediates. 
In \cite{pretsaICPM2019}, the authors apply $k$-anonymity and $t$-closeness \cite{li2007t} on event data to preserve the privacy of \textit{resources}. In \cite{MannhardtKBWM19}, the authors employ the notion of differential privacy \cite{Dwork08differential} to preserve the privacy of \textit{cases} in event logs. In \cite{rafieitlkc}, an efficient algorithm is introduced applying $k$-anonymity and \textit{confidence bounding} to preserve the privacy of \textit{cases} in event logs.

Most related to our work are \cite{PikaWBHAR19} and \cite{pika2020privacy} which are focused on healthcare event data. In \cite{PikaWBHAR19}, the authors analyze data privacy and utility requirements for healthcare event data and the suitability of privacy preservation techniques is assessed. In \cite{pika2020privacy}, the authors extend their previous research (\cite{PikaWBHAR19}) to demonstrate the effect of applying some anonymization operations on various process mining results, privacy metadata are advocated, and a privacy extension for the XES standard is proposed. 
However, formal models, possible risks, the tools support, and the event data which are not in the form of event logs are not discussed. In this paper, we provide a comprehensive infrastructure for recording privacy metadata which considers possible risks for data leakage, and the data utility. Our infrastructure is enriched by the formal models in theory and the tools support in practice.      

\section{Preliminaries}\label{sec:prelimineries}

In this section, we provide formal definitions for event logs used in the remainder.
Let $A$ be a set. $A^*$ is the set of all finite sequences over $A$, and $\multiset(A)$ is the set of all multisets over the set $A$. A finite sequence over $A$ of length $n$ is a mapping $\sigma \in \{1,...,n\} \rightarrow{A}$, represented as $\sigma = \langle a_1,a_2,...,a_n \rangle$ where $\sigma_i = a_i = \sigma(i)$ for any $1\leq i \leq n$, and $|\sigma|=n$. $a \in \sigma \Leftrightarrow{a=a_i}$ for $1 \leq i \leq n$. 
For $\sigma \in A^*$, $\{a \in \sigma\}$ is the set of elements in $\sigma$, and $[a \in \sigma]$ is the multiset of elements in $\sigma$, e.g., $[a \in \langle x,y,z,x,y \rangle ] = [x^2,y^2,z]$. $\sigma \downarrow_{X}$ is the projection of $\sigma$ onto some subset $X \subseteq A$, e.g., $\langle a,b,c \rangle \downarrow_{\{a,c\}} = \langle a,c \rangle $. 
$\sigma \cdot \sigma'$ appends sequence $\sigma'$ to $\sigma$ resulting a sequence of length $|\sigma| + |\sigma'|$.


\begin{definition}[Event, Attribute]
	\label{def:eventattribute}
	Let $\eventuniverse$ be the event universe, i.e., the set of all possible event identifiers. Events can be characterized by various attributes, e.g., an event may have a timestamp and corresponds to an activity. 
	Let $\attuniverse_{event}$ be the set of all possible event attribute names. For any $e \in \eventuniverse$ and name $n \in \attuniverse_{event}$, $\#_n(e)$ is the value of attribute $n$ for event $e$. For an event $e$, if $e$ does not have an attribute named $n$, then $\#_n(e)=\bot~(null)$.
\end{definition}


We assume three standard explicit attributes for events: \textit{activity}, \textit{time}, and \textit{resource}. We denote $\universe_{act}$, $\universe_{time}$, and $\universe_{res}$ as the universe of activities, timestamps, and resources, respectively. $\#_{act}(e) \in \universe_{act}$ is the \textit{activity} associated to event $e$, $\#_{time}(e) \in \universe_{time}$ is the \textit{timestamp} of event $e$, and $\#_{res}(e) \in \universe_{res}$ is the \textit{resource} associated to event $e$.


\begin{definition}[Case, Trace]
	\label{def:casetrace}
	Let $\caseuniverse$ be the case universe, i.e., the set of all possible case identifiers, and $\attuniverse_{case}$ be the set of case attribute names. For any case $c \in \caseuniverse$ and name $n \in \attuniverse_{case}{:}~ \#_n(c)$ is the value of attribute $n$ for case $c$. For a case $c$, if $c$ does not have an attribute named $n$, then $\#_n(c)=\bot~(null)$. Each case has a mandatory attribute \enquote{trace} such that $\#_{trace}(c) \in \eventuniverse^*$. A trace $\sigma$ is a finite sequence of events such that each event appears at most once, i.e., for $1 \leq i < j \leq |\sigma|: \sigma(i) \neq \sigma(j)$. 
\end{definition}

\begin{definition}[Event Log]
	\label{def:eventlog}
	Let $\valueuniverse$ be a universe of values including a designated null value $(\bot)$, and $\attuniverse = \attuniverse_{event} \cup \attuniverse_{case}$ such that $\attuniverse_{event}$ and $\attuniverse_{case}$ are disjoint. An event log is a tuple $L= (C,E,N,\#)$, in which;  $C \subseteq \caseuniverse$ is a set of case identifiers, $E \subseteq \eventuniverse$ is a set of event identifiers such that $E=\{ e \in \eventuniverse \mid \exists_{c \in C}~{e \in \#_{trace}(c)} \}$, $N \subseteq \attuniverse$ is a set of attribute names such that $N \cap \attuniverse_{event}$ are the event attributes, and $N \cap \attuniverse_{case}$ are the case attributes. For $n \in N$, $\#_n \in (C \cup E) \rightarrow{\valueuniverse}$ is a function which retrieves the value of attribute $n$ assigned to a case or an event $($$\bot$ is used if an attribute is undefined for the given case or event$)$.  
	In an event log, each event appears at most once in the entire event log, i.e., for any $c_1,c_2 \in C$ such that $c_1 \neq c_2: \{ e \in \#_{trace}(c_1) \} \cap \{ e \in \#_{trace}(c_2) \} = \emptyset$. If an event log contains timestamps, then the ordering in a trace should respect the timestamps, i.e., for any $c \in C$, $i$ and $j$ such that $1 \leq i < j \leq |\#_{trace}(c)|: \#_{time}(\#_{trace}(c)_i) \leq \#_{time}(\#_{trace}(c)_j)$.  We denote $\universe_{L}$ as the universe of event logs.
\end{definition}


\section{Privacy-Preserving Data Publishing}\label{sec:ppdp}

Privacy-preserving data publishing is the process of changing the original data before publishing such that the published data remain practically useful while individual privacy is protected \cite{fung2010introduction}. Note that the assumption is that the data analysts (process miners) are not trusted and may attempt to identify sensitive personal information. Various privacy models could be applied to the original data to provide the desired privacy requirements before data publishing. However, the transformations applied should be known in order to interpret the data. Therefore, we propose the privacy metadata in process mining based on the main anonymization operations. We consider \textit{suppression} ($sup$), \textit{addition} ($add$), \textit{substitution} ($sub$), \textit{condensation} ($con$), \textit{swapping} ($swa$), \textit{generalization} ($gen$), and \textit{cryptography} ($cry$) as the main anonymization operation types in process mining. In this section, we define these operations and demonstrate how the original event log is modified by them.

\subsection{Anonymization Operations}\label{anonymizagion_operations}

Anonymization operations are the main functions which modify the original event log to provide the desired privacy requirements and may have any number of parameters. Moreover, the operations can be applied at the case or event level, and the target of the operation could be a case, an event, or attributes of such an object. We define the \textit{anonymizer} as a function which applies an anonymization operation to an event log.


\begin{definition}[Anonymizer]
	\label{def:anonymizer}
	An anonymizer $anon \in \universe_{L} \rightarrow{\universe_{L}}$ is a function mapping an event log into another one by applying an anonymization operation.
\end{definition}

\begin{definition}[Anonymizer Signature]
	\label{def:anonymizer_sign}
	Let $OT = \{sup, add, sub, con,swa,\\ gen, cry \}$ be the set of operation types. A signature $sign = (ot,level,target) \in OT \times \{case,event\} \times (\{case,event\} \cup \attuniverse)$ characterizes an anonymizer by its type, the level $($case or event$)$, and the target $($case, event, or case/event attribute$)$. We denote $\universe_{sign}$ as the universe of signatures. 
\end{definition}

Note that an anonymizer signature is not supposed to uniquely distinguish the anonymization operations applied to an event log, i.e., the same type of operation can be applied at the same level and to the same target multiple times, but it is designated to reflect the \textit{type} and the \textit{direct purpose} of the corresponding operation. This is considered as the minimum information required to make the analysts aware of the modifications w.r.t. the data utility and risks for data leakage. We say the \textit{direct purpose} due to the interdependency of cases and events through traces, i.e., modifying cases may affect events and vice versa. This is demonstrated by some of the examples in the following.
We assume that the only correlation between cases and events is the trace attribute of cases, and all the other attributes are independent.
Certainly, the operations can be more accurately characterized by adding more information to the corresponding signatures, but this may lead to the data leakage in the sense of revealing the protected information or details of the anonymization operation which is orthogonal to the ultimate goal of privacy preservation.  
In the following, we demonstrate the anonymization operations using some examples.


\subsubsection{Suppression} replaces some values, specified by the target of the operation and probably some conditions, with a special value, e.g., null value. The reverse operation of suppression is called \textit{disclosure}. The group-based privacy models such as $k$-anonymity and its extensions \cite{li2007t} often utilize suppression. In the following, we provide three examples to demonstrate the effect of suppression on an event log. In \cite{pretsaICPM2019}, a group-based privacy model is proposed for discovering processes while preserving privacy.
 
	\begin{sloppy}
	
	\begin{exmp}\label{supEvent}\textnormal{\textbf{(Event-event suppression based on the activity attribute)}}
	\label{suppression1}
		Let  $L= (C,E,N,\#)$ be an event log. We want to suppress the events $e \in E$, if their activity attribute value is $a \in \pactivities$.
		$anon^{a}_{\ref{supEvent}}(L) = L'$ such that:
		$L'=(C', E', N',\#')$, $E'=\{ e \in E \mid \#_{act}(e) \neq a \}$, $C'= C$, and $N'=N$.
		$\forall_{e \in E'}\forall_{n \in N'}\#'_n(e)=\#_n(e)$,
		$\forall_{c \in C'}\forall_{n \in N'\setminus\{trace\}}\#'_n(c)=\#_n(c)$, and $\forall_{c \in C'}\#'_{trace}(c)=\#_{trace}(c) \downarrow_{E'}$.
		The anonymizer signature of this operation is $sign = (sup,event,event)$.
	\end{exmp}
	\end{sloppy}

	\begin{exmp}\label{supCase}\textnormal{\textbf{(Case-case suppression based on the trace length)}}\\
		Let $L= (C,E,N,\#)$ be an event log. We want to suppress the cases $c \in C$, if the trace length of the case is $k \in \mathbb{N}_{\geq 1}$.
		$anon^{k}_{\ref{supCase}}(L) = L'$ such that:
		$L'=(C', E', N',\#')$, $C'=\{ c \in C \mid |\#_{trace}(c)| \neq k \}$, $E' = \{e \in \eventuniverse \mid \exists_{c \in C'}e \in \#_{trace}(c) \}$, and $N'=N$.
		$\forall_{e \in E'}\forall_{n \in N'}\#'_n(e)=\#_n(e)$ and
		$\forall_{c \in C'}\forall_{n \in N'}\#'_n(c)=\#_n(c)$. 
		The corresponding signature for this operation is $sign=(sup,case,case)$.
	\end{exmp}

	\begin{exmp}\label{supRes}\textnormal{\textbf{(Event-resource suppression based on the activity attribute)}}\\
	Let $L= (C,E,N,\#)$ be an event log. We want to suppress the resource attribute value of the events $e \in E$, if their activity attribute value is $a \in \pactivities$. 
	$anon^{a}_{\ref{supRes}}(L) = L'$ such that:
	$L'=(C', E', N',\#')$, $C'= C$, $E'= E$, and $N'=N$. $\forall_{e \in E'}\forall_{n \in N'\setminus\{res\}}\#'_n(e)=\#_n(e)$.
	For all $e \in E'$: $\#'_{res}(e)=\bot$ if $\#_{act}(e) = a$, otherwise $\#'_{res}(e)=\#_{res}(e)$.
	$\forall_{c \in C'}\forall_{n \in N'}\#'_n(c)=\#_n(c)$.
	The corresponding anonymizer signature is $sign=(sup,event,resource)$.
	\end{exmp}


As above-mentioned, modifying cases may affect events and vice versa. In Example~\ref{supEvent}, event suppression modifies the cases through the trace attribute, which is an indirect effect not shown by the signature. Similarly, in Example~\ref{supCase}, suppressing cases results in event suppression, i.e., all the events in the trace of the suppressed cases are removed. This is also an indirect effect which is not reflected in the corresponding signature.

\subsubsection{Addition} is often used by the \textit{noise addition} techniques where the noise which is drawn from some distribution is randomly added to the original data to protect the sensitive values. In process mining, the noise could be added to cases, events, or the attribute values. In \cite{MannhardtKBWM19}, the notion of \textit{differential privacy} is used as a noise addition technique to perform privacy-aware process discovery. 
	
	\begin{exmp}\label{addTrace}\textnormal{\textbf{(Add an event to the end of traces based on the activity attribute of the last event in the trace)}}\\
		Let  $L= (C,E,N,\#)$ be an event log, and $N=\{trace,act,res,time\}$ such that \enquote{trace} is the case attribute and \enquote{act}, \enquote{res}, and \enquote{time} are the event attributes.
		We want to add an event $e' \in \eventuniverse \setminus E$ with the activity attribute value $a_2 \in \pactivities$ and the resource attribute value $r \in \presources$ at the end of the trace of a case $c \in C$, if the activity of the last event in the trace is $a_1 \in \pactivities$.
		$anon^{a_1,a_2,r}_{\ref{addTrace}}(L) = L'$ such that:
		$L'=(C', E', N',\#')$, $C'= C$, $N'=N$, $C_{cond}=\{c \in C \mid \#_{act}(\#_{trace}(c)_{|\#_{trace}(c)|})=a_1 \}$, and $f \in C_{cond} \rightarrow{\eventuniverse \setminus E}$ is a total injective function which randomly assigns unique event identifiers to the cases having the desired condition. 
		We denote $E_{add}=range(f)$ as the set of added events. Hence, $E'= E \cup E_{add}$.
		$\forall_{e \in E}\forall_{n \in N' \setminus \{trace\}}\#'_{n}(e)=\#_{n}(e)$.
		For all $c \in C_{cond}$: $\#'_{act}(f(c))=a_2$, $\#'_{time}(f(c))=\#_{time}(\#_{trace}(c)_{|\#_{trace}(c)|})+1$, and $\#'_{res}(f(c))=r$. 
		$\forall_{c \in C' \setminus C_{cond}}\#'_{trace}(c)=\#_{trace}(c)$, and $\forall_{c \in C_{cond}}\#'_{trace}(c)=\#_{trace}(c) \cdot \langle f(c) \rangle$.
		The corresponding signature for this operation is $sign=(add,case,trace)$.
	\end{exmp}

\subsubsection{Substitution} replaces some values with some substitutions specified by a set, i.e., a set of substitutions. The substitution could be done randomly or could follow some rules, e.g., a \textit{round robin} manner. In \cite{rafiei2019role}, a substitution technique is used in order to mine roles from event logs while preserving privacy. 	
	
	\begin{exmp}\label{subAct}\textnormal{\textbf{(Event-activity substitution based on a set of sensitive activities and a set of activity substitutions)}}\\
		Let  $L= (C,E,N,\#)$ be an event log, $A_L=\{a \in \pactivities \mid \exists_{e \in E} \#_{act}(e)=a \}$ be the set of activities in L, $A_x \subseteq \pactivities \setminus A_L$ be the set of activities used as the substitutions, $rand(X) \in X$ be a function which randomly returns an element from the set $X$, and $A_s \subset A_L$ be the set of sensitive activities. We want to randomly substitute the activity attribute value of the events $e \in E$, if the activity is sensitive.
		$anon^{A_s,A_x}_{\ref{subAct}}(L) = L'$ such that:
		$L'=(C', E', N',\#')$, $C'= C$, $E'= E$, and $N'=N$.
		For all $e \in E'$: $\#'_{act}(e)=rand(A_x)$ if $\#_{act}(e) \in A_s$, otherwise $\#'_{act}(e)=\#_{act}(e)$. 
		$\forall_{e \in E'}\forall_{n \in N' \setminus \{act\}}\#'_n(e)=\#_n(e)$, $\forall_{c \in C'}\forall_{n \in N'}\#'_n(c)=\#_n(c)$.
		$sign=(sub,event,activity)$ is the signature corresponds to this operation.
	\end{exmp}

\subsubsection{Condensation} first condenses the cases into similar clusters based on the sensitive attribute values, then in each cluster the sensitive attribute values are replaced with a collective statistical value, e.g., mean, mode, median, etc, of the cluster. In \cite{AggarwalY08}, the authors introduce condensation-based methods for privacy-preserving data mining. The following example shows how the condensation operates on the event logs assuming that there exists a sensitive case attribute.

	\begin{exmp}\label{consAtt}\textnormal{\textbf{(Case-attribute condensation based on a set of case clusters, a cluster finder function, and using mode as the collective value)}}\\
		Let  $L= (C,E,N,\#)$ be an event log, $CL=\{cl_1,cl_2,...,cl_n\}$ be the set of case clusters, whose sensitive attribute value is similar, such that for $1 \leq i \leq n$: $cl_i \subseteq C$ and for $1 \leq i,j \leq n$, $i \neq j$: $cl_i \cap cl_j = \emptyset$. 
		Also, let $f \in C \rightarrow{CL}$ be a function which retrieves the cluster of a case, and $n' \in N$ be the sensitive case attribute, e.g., disease. 
		For a set $X$, $mode(X) \in X$ retrieves the mode of the set. 
		We want to replace the value of $n'$ for each case $c \in C$ with the mode of $n'$ in the cluster of the case.  
		$anon^{CL,f,n'}_{\ref{consAtt}}(L) = L'$ such that:
		$L'=(C', E', N',\#')$, $C'= C$, $N'=N$, $E'= E$, 
		$\forall_{c \in C'}\#'_{n'}(c)=mode(\{ \#_{n'}(c') \mid c' \in f(c) \})$,
		$\forall_{c \in C'}\forall_{n \in N'\setminus\{n'\}}\#'_n(c)=\#_n(c)$, and $\forall_{e \in E'}\forall_{n \in N'}\#'_n(e)=\#_n(e)$.
		$sign=(con,case,n')$ is the signature corresponds to this operation.
	\end{exmp}

\subsubsection{Swapping} aims to anonymize data by exchanging values of a sensitive attribute between individual cases. The individual cases which are chosen to exchange the sensitive attribute values are supposed to have similar sensitive attribute values. Therefore, cases need to be clustered into the clusters with the similar sensitive attribute values. The cases for swapping in the same cluster could be chosen either randomly or by some methods, e.g., the \textit{rank swapping} method \cite{rankSwapping}. The following example shows how the swapping operates on the event logs assuming that there exists a sensitive case attribute.

	\begin{exmp}\label{swapAtt}\textnormal{\textbf{(Case-attribute swapping based on a set of case clusters and a cluster finder function)}}\\
		Let  $L= (C,E,N,\#)$ be an event log, $CL=\{cl_1,cl_2,...,cl_n\}$ be the set of case clusters, whose sensitive attribute value is similar, such that for $1 \leq i \leq n$: $cl_i \subseteq C$ and for $1 \leq i,j \leq n$, $i \neq j$: $cl_i \cap cl_j = \emptyset$. 
		Also, let $f \in C \rightarrow{CL}$ be the function which retrieves the cluster of a case, and $n' \in N$ be the sensitive case attribute, e.g., disease. 
		For a set $X$, $rand(X) \in X$ is a function which randomly retrieves an element from the set. 
		We want to randomly swap the value of $n'$ for each case $c \in C$ with the $n'$ value of another case in the same cluster.  
		$anon^{CL,f,n'}_{\ref{swapAtt}}(L) = L'$ such that:
		$L'=(C', E', N',\#')$, $C'= C$, $N'=N$, $E'= E$, 
		$\forall_{c \in C'}\#'_{n'}(c)=rand(\{ \#_{n'}(c') \mid c' \in f(c) \setminus \{c\} \})$,
		$\forall_{c \in C'}\forall_{n \in N'\setminus\{n'\}}\#'_n(c)=\#_n(c)$, and $\forall_{e \in E'}\forall_{n \in N'}\#'_n(e)=\#_n(e)$.
		The anonymizer signature of this operation is $sign=(swa,case,n')$.
	\end{exmp}

\subsubsection{Cryptography} includes a wide range of techniques such as \textit{encryption}, \textit{hashing}, \textit{encoding}, etc. In \cite{rafieiWA18}, the \textit{connector method} is introduced as an encryption-based technique which breaks the relation between the events in the traces, while discovering the directly follows graph (DFG) \cite{leemans2018scalable} from an event log. In the following example, we demonstrate the effect of applying an encryption technique on the activity attribute of the events in an event log.


	\begin{exmp}\label{encAct}\textnormal{\textbf{(Event-activity encryption based on an encryption method)}}\\
	Let $L= (C,E,N,\#)$ be an event log, $ENC$ be the universe of encryption method names, and $KEY$ be the universe of keys. We want to encrypt the activity attribute of all the events $e \in E$ using a method $m \in ENC$ and a key $k \in KEY$.
	Let $\valueuniverse$ be a universe of values, and $\valueuniverse_{enc}$ be a universe of encrypted values. $enc \in \valueuniverse \times ENC \times KEY \nrightarrow{\valueuniverse_{enc}}$ is a partial function which encryptes a value $u \in \valueuniverse$ given the name of method and a key.
	Given $k \in KEY$, and $m \in ENC$, $anon^{k,m}_{\ref{encAct}}(L) = L'$ such that:
	$L'=(C', E', N',\#')$, $C'= C$, $E'= E$, and $N'=N$.
	$\forall_{e \in E'}\#'_{act}(e)=enc(\#_{act}(e),m,k)$, 
	$\forall_{e \in E'}\forall_{n \in N' \setminus \{act\}}\#'_n(e)=\#_n(e)$, and
	$\forall_{c \in C'}\forall_{n \in N'}\#'_n(c)=\#_n(c)$.
	The signature of this operation is $sign=(cry,event,activity)$.
	\end{exmp}

\subsubsection{Generalization} replaces some values, indicated by the target of operation and probably some conditions, with a parent value in the taxonomy tree of an attribute. The reverse operation of generalization is called \textit{specialization}. The four main generalization schemes are \textit{full-domain}, \textit{subtree}, \textit{sibling}, \textit{cell} \cite{fung2010introduction}. 
In the \textit{full-domain} scheme, all values in an attribute are generalized to the same level of the taxonomy tree. 
In the \textit{subtree} scheme, at a non-leaf node, either all child values or none are generalized.
The \textit{sibling} generalization is similar to the \textit{subtree}, but some siblings may remain ungeneralized. In the \textit{cell} generalization, some instances of a value may remain ungeneralized while in all the other schemes if a value is generalized, all its instances are generalized.

The generalization techniques are often used by group-based anonymization techniques. In the following, we demonstrate the effect of applying the generalization operation on the event logs by a simple example that uses the \textit{full-domain} scheme to generalize the time attribute of the events.


	\begin{exmp}\label{genTime}\textnormal{\textbf{(Event-time generalization based on a time generalization method)}}\\
		Let $L= (C,E,N,\#)$ be an event log and $TL= \{seconds, minutes, hours, days,\\ months, years\}$ be the level of time generalization. $g_{tl} \in \ptimes \rightarrow{\ptimes} $ is a function that generalizes timestamps to the given level of time. 
		We want to generalize the time attribute of all the events $e \in E$ to the level $tl \in TL$. 
		$anon^{tl}_{\ref{genTime}}(L) = L'$ such that:
		$L'=(C', E', N',\#')$, $C'= C$, $E'= E$, and $N'=N$. 
		$\forall_{e \in E'}\#'_{time}(e)=g_{tl}(\#_{time}(e))$, 
		$\forall_{e \in E'}\forall_{n \in N' \setminus \{time\}}\#'_n(e)=\#_n(e)$,
		$\forall_{c \in C'}\forall_{n \in N'}\#'_n(c)=\#_n(c)$.
		$sign=(gen,event,time)$ is the anonymizer signature of this operation.
	\end{exmp}
	
%

Privacy-preserving data publishing in process mining, is done by applying a sequence of the anonymization operations to fulfill a desired privacy requirement. Note that the operations are often applied by the privacy models, where the data utility preservation is also taken into account. 
  
\begin{sloppy}
	\begin{definition}[Privacy Preserving Data Publishing in Process Mining - $ppdp$]
		\label{def:ppdp}
		Let $pr$ be the desired privacy requirement and $\universe_{L}$ be the universe of event logs, $ppdp^{pr}: \universe_{L} \rightarrow{\universe_{L}}$ is a privacy method that gets an event log and applies $i \in \mathbb{N}_{\geq 1}$ anonymization operations to the event log in order to provide an anonymized event log which satisfies the given privacy requirement $pr$. If we assume that $pr$ is satisfied at the step (layer) $n$ of the anonymization process, then for $1 \leq i \leq n$, $L_i=anon_i(L_{i-1})$ such that $L_0 = L$ and $L_n$ is the anonymized event log which satisfies $pr$. 
	\end{definition}
\end{sloppy}


\subsection{Data Utility and Data Leakage}\label{sec:utility_leakage}
We define \textit{potential original event logs} to show how the anonymizer signature can be exploited to narrow down the set of possible original event logs.

\begin{definition}[Potential Original Event Log]
	\label{def:pol}
	Let $\universe_{L}$ be the universe of event logs and $\universe_{sign}$ be the universe of signatures. $po \in \universe_{L} \times \universe_{L} \times \universe_{sign} \rightarrow{\mathbb{B}} $ is a function that for a given event log, an anonymized event log, and the signature of the anonymized event log checks whether the given event log could be an original event log. $ol \in \universe_{L} \times \universe_{sign} \rightarrow{2^{\universe_{L}}}$ retrieves a set of potential original event logs, s.t., for $L' \in \universe_{L}$ and $sign \in \universe_{sign}$: $ol(L',sign)=\{ L \in \universe_{L} \mid po(L,L',sign) \}$.
\end{definition}

To demonstrate the effect of the anonymizer signature, as a privacy metadata candidate, on \textit{data utility} and \textit{data leakage}, we analyze the set of potential original event logs. Here, the analysis is performed for Example~\ref{supEvent}. However, the concept is general and can be applied to all the operations.
Let $L'= (C',E',N',\#')$ be an anonymized event log and $sign=(sup,event,resource)$ . 
For an event log $L=(C,E,N,\#)$: $po(L,L',sign) = true$ $\iff$
$(C=C' \wedge E = E' \wedge N=N' \wedge
\forall_{e \in E'}\forall_{n \in N'\setminus \{res\}}\#_n(e)=\#'_n(e) \wedge
\forall_{\{e \in E' \mid \#'_{res}(e) = \bot\}}\#_{res}(e) \in \universe_{res} \wedge 
\forall_{\{e \in E' \mid \#'_{res}(e) \ne \bot\}}\#_{res}(e) = \#'_{res}(e) \wedge
\forall_{c \in C'}\forall_{n \in N'}\#_n(c)=\#'_n(c))$.
$ol(L',sign)=\{ L \in \universe_{L} \mid po(L,L',sign) \}$ is the set of potential original event logs. 
Intuitively, $|ol(L',sign)|=|\{ e \in E' | \#'_{res}(e)=\bot \}| \times |\universe_{res}|$, where $\universe_{res}$ can be narrowed down to a few resources based on some background knowledge. 

If we ignore the target information in the signature, i.e., $sign=(sup,event)$, the uncertainty regarding the original event log and the results of analyses would be much higher, since the suppressed data could be some events, or any event attribute $n$, s.t., $\exists_{e \in E}\#'_n(e)=\bot$. That is, the uncertainty regarding the results of analyses is expanded from the \text{resource perspective} to all the perspectives. 
In contrast, if we add more information to the signature, reconstructing the original event log could be easier for an adversary. For example, if we add the condition of resource suppression to the signature, i.e., $sign=(sup,event, resource, (activity=a))$, then $|ol(L',sign)|=|\{ e \in E' | (\#'_{res}(e)=\bot \wedge \#'_{act}(e)=a) \}| \times |\universe_{res}|$. That is, the only information that an adversary needs to reconstruct the original event log, with high confidence, is to realize the set of resources who perform the activity $a$.
These analyses demonstrate that privacy metadata should contain the minimum information that preserves a balance between data utility and data leakage. 

\subsection{Privacy Metadata}\label{sec:privacy_metadata}

\begin{figure}[t]
	\centering
	\includegraphics[width=0.90\textwidth]{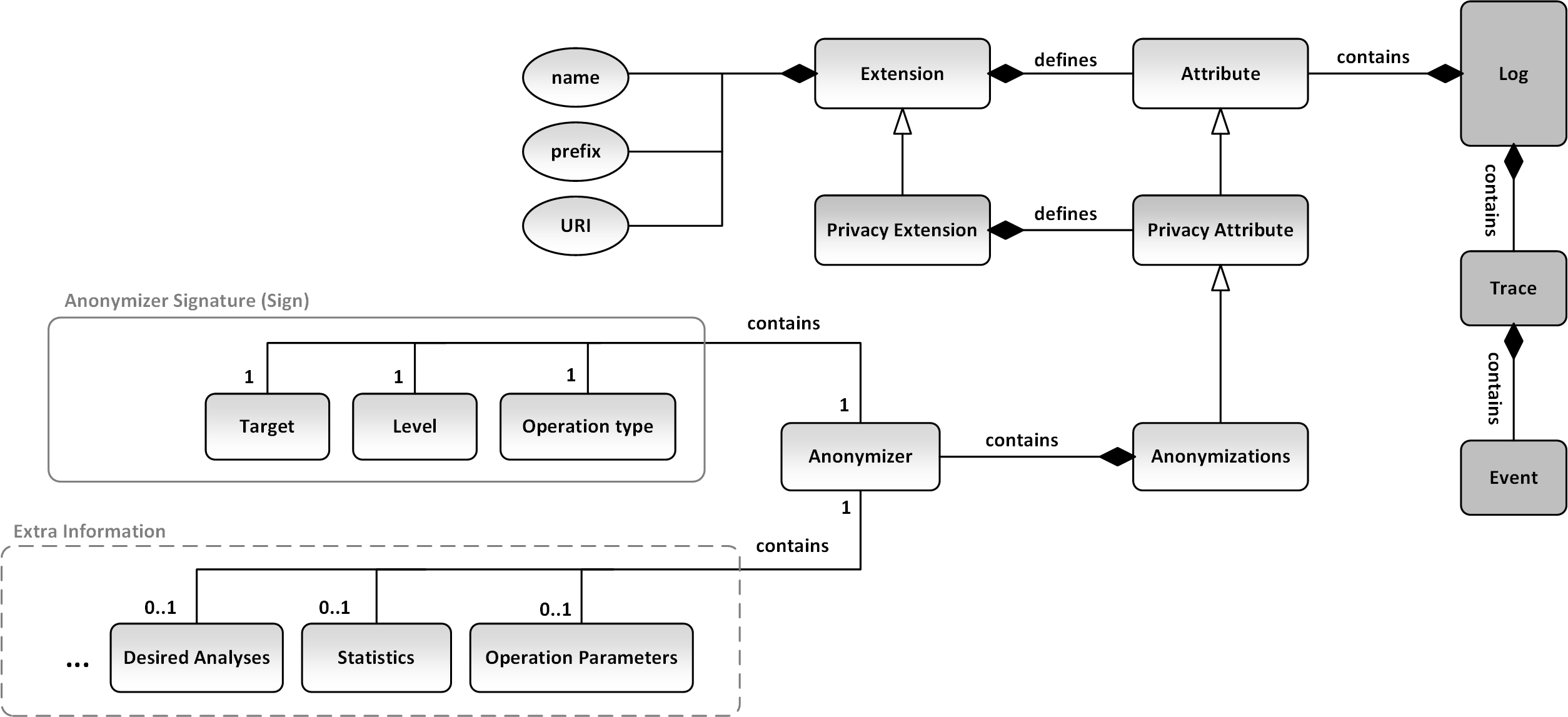}
	\caption{The meta model of the proposed privacy metadata which is presented as an extension in XES. The privacy metadata attributes are added as the log attributes.}\label{fig:extension_concept}
\end{figure}

Privacy metadata in process mining should correspond to the privacy-preserving data publishing and are supposed to capture and reflect the modifications, made by the anonymization operations. For event data in the form of an event log, privacy metadata are established by the means of XES. 
\autoref{fig:extension_concept} shows the meta model of the proposed privacy metadata as an extension in the XES. The privacy metadata attributes are defined by an extension at the \textit{log level} . 
Note that the level (log, trace, or event) that is chosen to include the privacy metadata is one of the important risk factors. Although the anonymization operations are applied at the case and event levels, adding the corresponding metadata to the same level is of high risk and may lead to the protected data leakage. 
Assume that we add the privacy metadata regarding the resource suppression operation applied to \autoref{tbl:sample_evenlog} at the event level. By doing so, we expose the exact event that has been affected, and consequently, all the other information about the event are exposed, e.g, the activity performed by the removed resource is \enquote{r}, and background knowledge could connect this information to the protected sensitive data. A similar risk scenario can be explored for the activity substitution operation applied to \autoref{tbl:sample_evenlog}. If the corresponding metadata is added at the event level, the substitution set of the activities could be partially or entirely exposed.

In \autoref{fig:extension_concept}, \textit{anonymizations} is the main privacy metadata element which contains at least one \textit{anonymizer} element. Each \textit{anonymizer} element corresponds to one anonymizer which is applied at a layer of the anonymization process in \autoref{def:ppdp}. The \textit{anonymizer} element contains two types of attributes: \textit{mandatory} and \textit{optional}. The mandatory attributes, residing in the solid box in \autoref{fig:extension_concept}, correspond to the \textit{anonymizer signature} (\autoref{def:anonymizer_sign}) and should be reflected in the XES. However, the optional attributes, residing in the dash box in \autoref{fig:extension_concept}, are the attributes which could be included in the XES. \textit{Operation parameters} could contain the parameters which are passed to an anonymization operation, e.g., in Example~\ref{encAct}, the name of encryption method is an operation parameter. \textit{Statistics} could contain some statistics regarding the modification, e.g., the number of modified events, and \textit{desired analyses} could contain the appropriate process mining activities which are applicable after modifying the event log. Note that the more unnecessary information included in the privacy metadata, the more risk is accepted. For example, if the \textit{statistics} is included in an XES file where it is indicated that only 10 out of 1000 events have been modified, then an adversary makes sure that the transformed event log is almost the same as the original one (99\% similarity).

\begin{figure}[tb]
	\centering
	\frame{\includegraphics[width=0.60\textwidth]{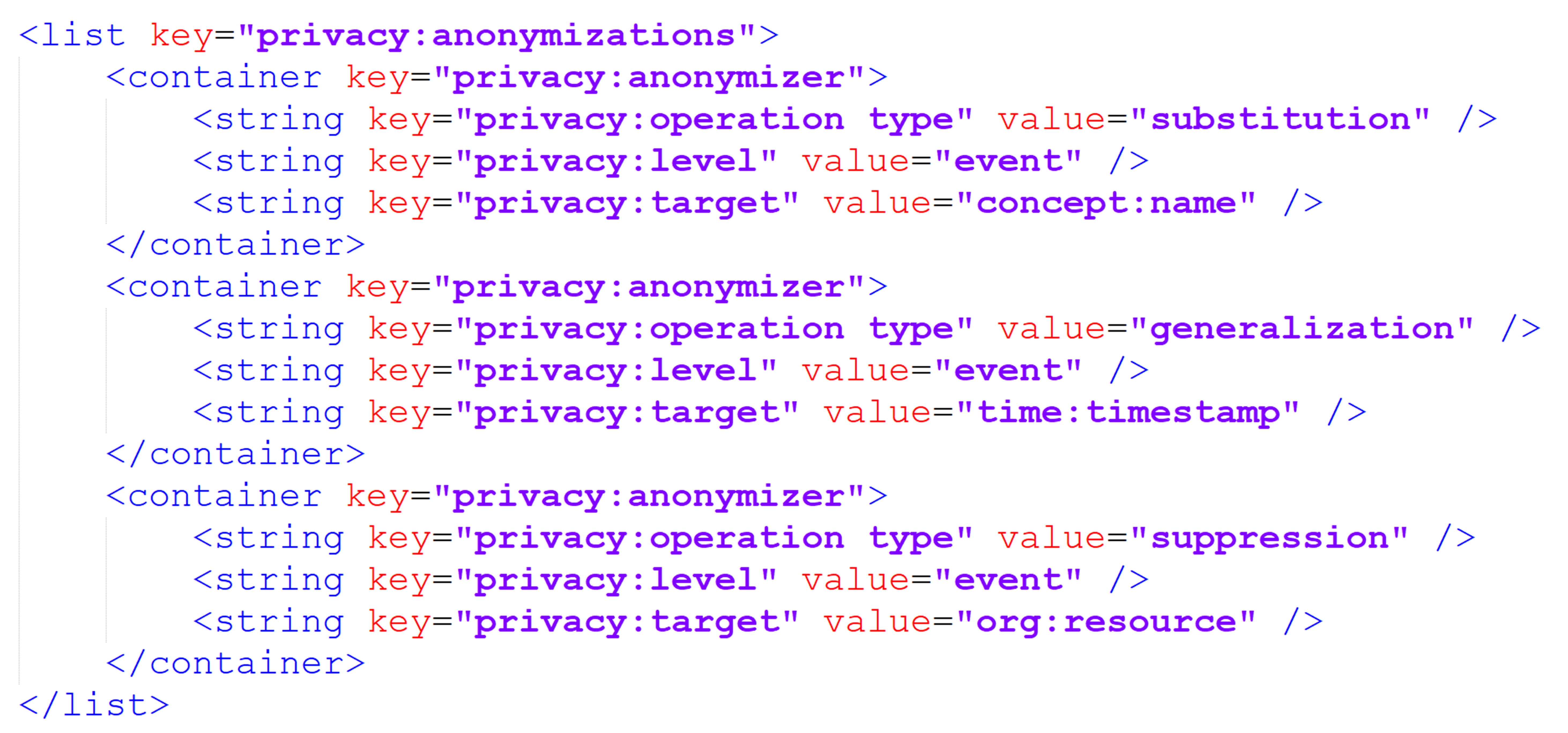}}
	\caption{Privacy metadata recorded after transforming  \autoref{tbl:sample_evenlog} to \autoref{tbl:sample_evenlog_anon} in order to satisfy a privacy requirement. \textit{privacy} is considered as the \textit{prefix} of the extension.}\label{fig:extension_xes}
\end{figure}

\autoref{fig:extension_xes} shows the privacy metadata recorded after transforming \autoref{tbl:sample_evenlog} to \autoref{tbl:sample_evenlog_anon}. Note that \textit{privacy} is considered as the \textit{prefix} of the extension. At the first layer of the anonymization, a \textit{substitution} operation has been applied which corresponds to Example~\autoref{subAct}, where $A_s=\{f\}$ and $A_x=\{g,k\}$. The metadata at this layer notices the analysts that some activity substitutions have been done at the event level, but it reveals neither the set of sensitive activities nor the substitution set. At the second layer of the anonymization, a \textit{generalization} operation has been done which generalizes the timestamps to the minutes level and corresponds to Example~\autoref{genTime}, where $tl=minutes$. At the last layer of the anonymization ($n=3$ in \autoref{def:ppdp}), a \textit{suppression} operation has been done which suppresses some resources and corresponds to Example~\autoref{supRes}, where the activity attribute value is $r$. This lets the analysts know that some resource suppression have been done without revealing the corresponding event. Note that the concept of \textit{layer} is implicitly established by the means of list, which imposes an order on the keys. 

\begin{figure}[bt]
	\centering
	\frame{\includegraphics[width=0.75\textwidth]{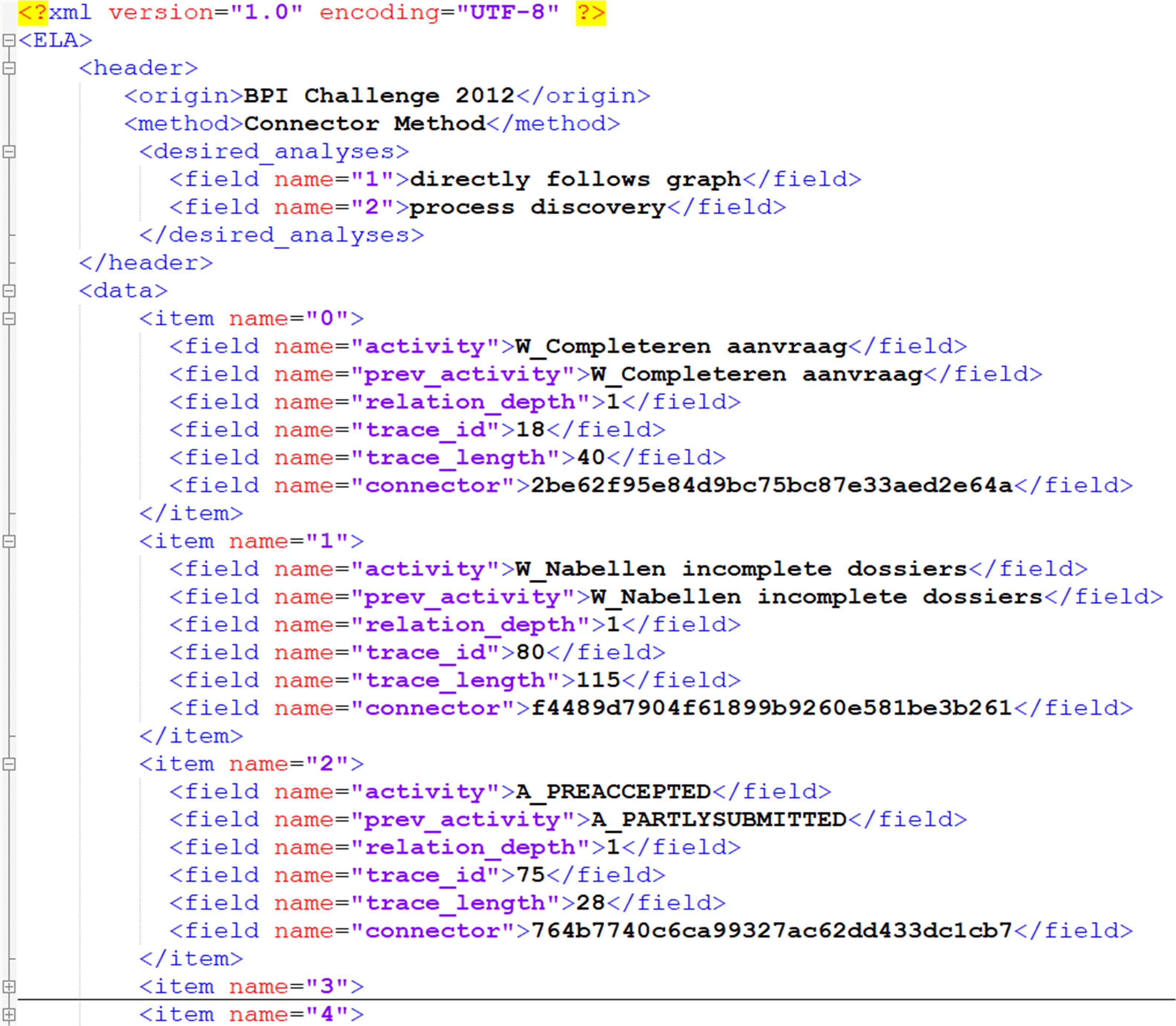}}
	\caption{The event log abstraction derived from the event log \enquote{BPI Challenge 2012} after applying the \textit{connector} method. Only the first 3 items are shown.}\label{fig:ela}
\end{figure}

So far, we only focused on the privacy preservation techniques applying the main anonymization operations on the original event log and transform it to another event log. However, there could be the techniques that do not preserve the structure of a standard event log. For example, in \cite{rafieiWA18}, the original event log is transformed to another form of event data where \textit{directly-follows relations} are extracted from the original traces for discovering \textit{directly-follows graph}. Such intermediate results derived from event logs and intended to relate logs and models are called \textit{abstractions} \cite{van2018process}. We introduce \textit{Event Log Abstraction} (ELA) to deal with the intermediate results created by some privacy preservation techniques which are not in the form of standard event logs. ELA is an XML tag-based language composed of two main parts: \textit{header} and \textit{data}. The \textit{header} part contains the privacy/confidentiality metadata, and the \textit{data} part contains the data derived from the original event log. The privacy metadata in ELA includes: \textit{origin}, \textit{method}, and \textit{desired analyses}. The \textit{origin} tag shows name of the event log the abstraction derived from, the \textit{method} tag contains name of the method, and the \textit{desired analyses} contains list of the appropriate analyses w.r.t. the abstraction. The \textit{data} part represents the data derived from the log in a tabular manner. \autoref{fig:ela} shows the ELA derived from the event log \enquote{BPI Challenge 2012} \cite{BPIC2012} by applying the \textit{connector} method \cite{rafieiWA18}.


\subsection{Tool Support}\label{sec:tool_support}
Since most of the privacy preservation techniques in process mining have been developed in Python, in order to support the tools, we have developed a Python library which is published as a Python package.\footnote{\scriptsize pip install p-privacy-metadata}\footnote{\scriptsize https://github.com/m4jidRafiei/privacy\_metadata} The library is based on $PM4Py$ \cite{berti2019process} and includes two classes correspond to two types of event data generated by the privacy preservation techniques in process mining: \textit{privacyExtension} and \textit{ELA}. The general advantage of the privacy metadata library is that the privacy experts do not need to deal with the tag-based files in order to read (write) the privacy metadata. 
Another crucial requirement provided by the \textit{privacyExtension} class is that it keeps the consistency of the privacy metadata by managing the order of the \textit{anonymizers} in the \textit{anonymizations} list. This class provides four main methods: \textit{set\_anonymizer}, \textit{set\_optional\_anonymizer}, \textit{get\_anonymizations}, and \textit{get\_anonymizer}. The \textit{set\_anonymizer} method gets the mandatory attributes and appends them to the \textit{anonymizations} list as an \textit{anonymizer} if there already exists a privacy extension, otherwise it first creates a privacy extension and an \textit{anonymizations} list, then adds the attributes to the list. The \textit{set\_optional\_anonymizer} method is responsible to add the optional attributes and should be called after setting the mandatory attributes. This method gets the layer, which is an index in the \textit{anonymizations} list, and optional attributes and adds the attributes to the given layer. The \textit{get\_anonymizations} method returns the whole \textit{anonymizations} tag in the XES file as a Python \textit{dictionary}. The \textit{get\_anonymizer} method gets a layer and returns the metadata of the layer.

\section{Conclusions}\label{sec:conclusions}
Due to the fact that event logs could contain highly sensitive personal information, and regarding the rules imposed by the regulations, e.g., GDPR, privacy preservation in process mining is recently receiving more attention. Event logs are modified by privacy preservation techniques, and the modifications may result in the event data which are not appropriate for all the process mining algorithms. In this paper, we discussed types of event data generated by the privacy preservation techniques.
We provided formal definitions for the main anonymization operations in process mining. Privacy metadata were proposed for event logs which are supported by formal definitions in order to demonstrate the completeness of the proposed infrastructure. 
The ELA (Event Log Abstraction) was introduced to cope with event data which are not in the form of standard event logs. 
We employed the IEEE XES standard in order to consistently develop our proposal for privacy metadata of event logs in practice, where a privacy extension was introduced. We also provided a Python library to support the privacy preservation tools for process mining which have been often developed in Python.

\section*{Acknowledgment} Funded under the Excellence Strategy of the Federal Government and the L{\"a}nder. We also thank the Alexander von Humboldt (AvH) Stiftung for supporting our research.

\bibliographystyle{splncs04}
\bibliography{Refrences}

\end{document}